\documentclass[letterpaper,twoside]{JHEP3}

\usepackage{amsmath}
\usepackage{mathrsfs}
\usepackage{graphicx}

\newcommand{\gev}{{\ \rm GeV}}
\newcommand{\tev}{{\ \rm TeV}}
\newcommand{\Tr}{{\rm Tr}}
\newcommand{\eg}{{\it e.g} }

\newskip\zatskip \zatskip=0pt plus0pt minus0pt
\def\matth{\mathsurround=0pt}

\def\atversim#1#2{\lower0.7ex\vbox{\baselineskip\zatskip\lineskip\zatskip
  \lineskiplimit 0pt\ialign{$\matth#1\hfil##\hfil$\crcr#2\crcr\sim\crcr}}}

\author{Ben Lillie\thanks{Work supported by the
  Department of Energy
  Contract DE-AC02-76SF00515.}
 \\ Stanford Linear Accelerator Center, 2575 Sand Hill
  Rd. Menlo Park, CA 94025\\
E-mail: \email{lillieb@slac.stanford.edu}}

\title{\center{Collider phenomenology of Higgs bosons in Left-Right symmetric Randall-Sundrum models}}
\preprint{SLAC-PUB-11176} 

\abstract{ We investigate the collider phenomenology of a left-right
symmetric Randall-Sundrum model with fermions and gauge bosons in
the bulk. We find that the model is allowed by precision electroweak
data as long as the ratio of the (unwarped) Higgs vev to the
curvature scale is $v/k \le 1/4$. In that region there can be
substantial modifications to the Higgs properties. In particular,
the couplings to $WW$ and $ZZ$ are reduced, the coupling to gluons
is enhanced, and the coupling to $\gamma\gamma$ can receive shifts
in either direction. The Higgs mass bound from LEP II data can
potentially be relaxed to $m_H \gtrsim 80$ GeV.}

\begin{document}

\section{Introduction}\label{sec:introduction}

While the Standard Model (SM) is a spectacularly successful
description of high energy particle phenomena, it leaves unexplained
why the Electroweak scale is much smaller than the GUT or Planck
scales. Recently, it has been proposed that this hierarchy might be
explained by the presence of additional compactified dimensions.
These could be TeV scale and flat
\cite{Arkani-Hamed:1998rs,Antoniadis:1990ew}, or Planck scale with a
warped geometry \cite{Randall:1999ee}. In this second scenario, the
Randall-Sundrum (RS) model, there is a single extra dimension and
the spacetime has the geometry of five-dimensional Anti-de Sitter
space, $AdS_5$, compactified on an orbifolded circle, $S^1/Z_2$. One
3-brane is localized on each end of the orbifold, and the warping
between them generates the Electroweak scale.

In the original RS model, all SM fields are localized on the TeV (or
IR) brane. The observable phenomenology in this case comes from the
new spin-2 graviton resonances \cite{Davoudiasl:1999jd}. However,
there is no reason for the fermions and gauge fields not to
propagate in the bulk
\cite{Davoudiasl:1999tf,Grossman:1999ra,Davoudiasl:2000wi,Pomarol:1999ad}.
Indeed, bulk fermions have a zero mode which is exponentially
localized near one of the branes. Choosing ${\mathcal O}(1)$
Lagrangian parameters for different fermions can select
exponentially different overlaps on the IR brane, and hence
different masses, providing a possible explanation of the fermion
mass hierarchy
\cite{Davoudiasl:2000wi,Gherghetta:2000qt,Hewett:2002fe}. There have
been many investigations into the phenomenology of this model
\cite{Csaki:2002gy,Carena:2002me,Carena:2002dz,Davoudiasl:2002ua,Huber:2003tu}.
One important conclusion is that simply putting the SM gauge group
in the bulk produces a large Peskin-Takeuchi $T$-parameter
\cite{Peskin:1990zt}. The can be fixed by expanding the gauge group
to be left-right symmetric, $SU(2)_L \times SU(2)_R \times
U(1)_{B-L}$ \cite{Agashe:2003zs} or by introducing brane localized
kinetic terms for the fermions \cite{Carena:2004zn}. It is also
possible to extend this to a Grand Unified group which then contains
a dark matter candidate \cite{Agashe:2004ci,Agashe:2004bm}.

More recently, there has been a proposal that no Higgs is needed in
this model, as the geometry can set the gauge boson masses
\cite{Csaki:2003dt,Csaki:2003zu,Nomura:2003du}, as well as the
fermion masses \cite{Csaki:2003sh}. The most straightforward
application of this model produces large electroweak corrections,
and does not preserve tree-level unitarity in longitudinal gauge
boson scattering at finite center-of-mass scattering energies
\cite{Barbieri:2003pr,Davoudiasl:2003zt,Ohl:2003dp,Davoudiasl:2003me,Davoudiasl:2004pw,Hewett:2004dv,Schwinn:2004xa,Cacciapaglia:2004jz,Hirn:2004ze}.
They do, however, contain a rich collider phenomenology which is
largely independent of those considerations \cite{Birkedal:2004au}.
There have been several variations on the model, including
deconstructing the extra dimension
\cite{Chivukula:2004af,Chivukula:2004pk,Chivukula:2005bn,Chivukula:2005xm,SekharChivukula:2004mu,Casalbuoni:2004id,Evans:2004rc},
adding additional dimensions \cite{Gabriel:2004ua}, and adding
additional branes \cite{Cacciapaglia:2005pa}.

The Higgsless models can be obtained as the limit of a model with a
Higgs where the vev, $v$, is taken to be large compared to the $AdS$
curvature $k$; that is $v/k \gg 1$. The fact that Higgsless
scenarios have difficulty accommodating the precision electroweak
observables leads to the speculation that the agreement may be
improved by including the effects of a finite Higgs vev.

Additionally, analysis of the Higgs scenario, $v/k \ll 1$ show that
this model is allowed by precision electroweak data. However, in
this case, with Kaluza-Klein (KK) masses, $m_{KK} \approx 2-3 \tev$,
we have $v/k \approx 1/4$, which is not particularly small. In fact,
this is expected from RS effective field theory arguments which tell
us that all Lagrangian level mass parameters should be of the same
order, $M_{\rm Pl}$. Hence, it makes sense to study the corrections
to collider observables induced by a finite, but not large, Higgs
vev.

In this paper we will examine the numerical behavior of explicit
solutions for the gauge and fermion wavefunctions in RS with a
finite and arbitrary value for $v/k$. From this we can extract both
the small vev limit and the Higgsless limit.

Note that the RS models can be thought of as large $N$ 4D conformal
gauge theories through the $AdS/CFT$ correspondence
\cite{Maldacena:1997re}. Analyses have also been performed on the
CFT side of the Higgsless model \cite{Burdman:2003ya}, and the Higgs
model \cite{Agashe:2004rs,Agashe:2005vg}. Note also that there is
potentially a term that mixes the Higgs field with the scalar
curvature, inducing mixing between the Higgs and radion states
\cite{Giudice:2000av,Csaki:2000zn}. However, the coefficient of this
term can be set to zero, which we will do here.

In Section \ref{sec:formalism} we develop the formalism that will be
employed in this paper. Section \ref{sec:spectra} shows the behavior
of the Kaluza-Klein spectra as the Higgs vev is varied. We
investigate the gauge couplings and precision electroweak
constraints in Section \ref{sec:gaugeff}, and the corrections to
Higgs properties in Section \ref{sec:higgs}. Section
\ref{sec:conclusion} concludes.

\section{Formalism}\label{sec:formalism}

We work in a slice of $AdS_5$, with metric (in conformal
coordinates)
\begin{gather}
ds^2 = \left(\frac{R}{z}\right)^2(d x^2-dz^2)
\end{gather}
where $R=1/k$ is the inverse of the curvature scale. There is one
brane located at $z=R$ (the Planck or UV brane), and a second brane
at $R'=(M_{\rm Pl}/TeV)R$ (the TeV or IR brane). This gives
$\log(R'/R) \sim 35$. We define $\epsilon = R/R' \sim 10^{-15}$ for
later convenience.

We are interested in models with a bulk $SU(2)_L\times SU(2)_R
\times U(1)_{B-L}$ gauge symmetry. The additional $SU(2)_R$ factor
over the SM provides a bulk custodial $SU(2)_{\rm c}$, which can
successfully protect the $T$ parameter \cite{Agashe:2003zs}. The
bulk gauge action we consider is then
\begin{gather}
 S_{bulk} = \int d^5x \sqrt{g} \left(
 \frac{-1}{4g^2_{5L}}F^L_{MN}F^{MN}_L
 +\frac{-1}{4g^2_{5R}}F^R_{MN}F^{MN}_R
 +\frac{-1}{4g^2_{5B}}F^B_{MN}F^{MN}_B
 \right)
\end{gather}
(Note that there is also a term for the gluon fields, which must
also be bulk fields. However, this term is unimportant for this
paper). We will make use of the ratios of gauge couplings
\begin{gather}
 \kappa = g_{5R}/g_{5L}, \qquad \lambda = g_{5B}/g_{5L}.
\end{gather}
Clearly, this group needs to be broken to $U(1)_{\rm EM}$. We
accomplish this by separating the breaking into two sectors. On the
Planck brane we break $SU(2)_R \times U(1)_{B-L} \rightarrow
U(1)_Y$. Since the UV brane is the only place where the $SU(2)_{\rm
c}$ is broken, we can see why the effects on the $T$ parameter will
be small. On the TeV brane we break $SU(2)_L \times SU(2)_R
\rightarrow SU(2)_D$, where $SU(2)_D$ is the diagonal subgroup of
$SU(2)_L$ and $SU(2)_R$.

We will work in the $A^5 = 0$ (unitary) gauge. The gauge condition
can potentially be complicated by the fact that the brane localized
Goldstone modes, the $G^i$ can mix with the $A^5$ modes. This means
that the physical longitudinal polarization for each vector is a
combination of bulk and brane modes. However, for the breaking
pattern used here there is no zero mode for any of the $A^5$ fields,
and hence no extra physical zero-mode scalar. We can therefore
safely work in the gauge where $A^5 = G^i = 0$.

We now ask what drives the breaking on each brane. On the Planck
brane all degrees of freedom will have Planck scale masses, so we
can ignore them. We can then implement the breaking with boundary
conditions to good approximation. This leads to the boundary
conditions at $z=R$
\begin{align}
 \partial_z \left(\frac{\kappa}{\lambda}A_R - A_B\right) & = 0, &\partial_z A_L & = 0,\hfill\notag \\
 A_B - \frac{\kappa}{\lambda}A_R^3 & = 0, & A_R^{\pm} & =
 0.\label{eq:gaugebcR}
\end{align}

On the TeV brane, the masses will be TeV scale, so we should look at
the Higgs sector in detail. The simplest structure that will create
the breaking pattern is a real Higgs that is a bidoublet under
$SU(2)_L \times SU(2)_R$. This leads to the boundary conditions at
$z=R'$
\begin{align}
 \partial_z (A_L + \kappa A_R) & = 0, & \partial_z A_B = 0,\hfill\notag\\
 \partial_z(\kappa A_L - A_R) & = -\frac{g^2_{5L}v^2}{4}(\kappa A_L - A_R).
 &\label{eq:gaugebcRp}
\end{align}
Note that in the $v/k\to\infty$ limit we obtain the usual Higgsless
boundary conditions. Instead of the real bidoublet, we could also
use the complex bidoublet familiar from Left-Right symmetric models
with minimal changes. See Appendix A for details.

To write down the effective 4D theory we expand the 5D fields into
Kaluza Klein (KK) fields.
\begin{gather}
A(x,z) = \sum_n \zeta^{(n)}_A(z)A^{(n)}(x)
\end{gather}
We can now obtain the gauge boson wavefunctions by solving the
equation of motion subject to the boundary conditions
(\ref{eq:gaugebcR}) and (\ref{eq:gaugebcRp}). The generic solution
for the wavefunctions is
\begin{gather}
\zeta_A^{(n)}(z) = z(A_A^{(n)}J_1(m_n z) + B_A^{(n)}Y_1(m_n
z)).\label{eq:gaugewf}
\end{gather}
Here the label $A$ refers to the particular gauge field being
expanded. One of the coefficients, $A^{(n)}$ and $B^{(n)}$, and the
mass are determined by inserting eq. (\ref{eq:gaugewf}) into eqs.
(\ref{eq:gaugebcR}) and (\ref{eq:gaugebcRp}). The other coefficient
is fixed by the normalization condition. We will use 4D canonical
normalization for all fields, giving
\begin{gather}
N_W^{(n)2} =  \int_R^{R'}\frac{dz}{z}R\left(
 |\zeta^{(n)}_{A^\pm_L}(z)|^2
 +|\zeta^{(n)}_{A^\pm_R}(z)|^2
 \right),
\end{gather}
for the charged gauge bosons, and
\begin{gather}
N_Z^{(n)2} =  \int_R^{R'}\frac{dz}{z}R\left(
 |\zeta^{(n)}_{A^0_L}(z)|^2
 +|\zeta^{(n)}_{A^0_R}(z)|^2
 +|\zeta^{(n)}_{B_R}(z)|^2
 \right),\label{eq:neutralnorm}
\end{gather}
for the neutral tower.

The fermion sector of the theory is more intricate. First, we will
need to arrange the SM fermions into representations of $SU(2)_R$.
There are two ways to do this in the RS model. The most
straightforward is to pair corresponding $SU(2)_L$ singlets into a
single $SU(2)_R$ doublet. So, \eg $u_R$ and $d_R$ become
$\begin{pmatrix} u_R & d_R \end{pmatrix}^{\top}$. The other option
is to make each right-handed field part of a different $SU(2)_R$
doublet. So $u_R \to \begin{pmatrix} u_R & d'_R
\end{pmatrix}^{\top}$, etc.. Orbifold projections are then required
to insure that there is no light mode for the new fermion states.
The first option follows more naturally from Grand Unified theories,
and allows the possibility of an explicit $Z_2$ symmetry that
exchanges the Left and Right gauge groups. The second makes it
easier to obtain top-bottom splitting and to suppress corrections to
the $Z b\bar b$ vertex, and is compatible with the GUT scenario in
\cite{Agashe:2004bm}. Here we will study the case where there is an
explicit $Z_2$, and hence choose to combine right-handed fields into
a single $SU(2)_R$ multiplet.

We write the 5D fermion as two 4D Majorana fermions, $\Psi_i =
(\psi_i \ \chi_i)^{\top}$. The orbifold conditions tell us that one
component must be even and the other odd \cite{Grossman:1999ra}. We
will pick the $\psi_i$ to be even for fields corresponding to the
left-hand SM fermions, and the $\chi_i$ even for the right-handed
ones. The Yukawa couplings to $\phi$, the Higgs on the IR brane,
will connect the left and right-handed zero modes and lift them.
These couplings are
\begin{gather}
 S_{IR} = \sum_f \int d^4 x \left(\frac{R}{z}\right)^4 \lambda_f \phi \left(
 \psi^f_{Ri}\chi^f_{Li} + \bar\chi^f_{Li}\bar\psi^f_{Ri}
 +\psi^f_{Li}\chi^f_{Ri} +
 \bar\chi^f_{Ri}\bar\psi^f_{Li}\right).\label{eq:yukawas}
\end{gather}
Here $f$ labels the fermion flavor. Eq. (\ref{eq:yukawas}) induces
the boundary conditions at $z=R'$
\begin{gather}
\psi^f_L = -\lambda_f v \psi^f_R \qquad \chi^f_R = \lambda_f v
\chi^f_L.\label{eq:fermionIR}
\end{gather}
This is equivalent to introducing a Dirac mass $\lambda_f v$ on the
IR brane. Note that this is $SU(2)_D$ symmetric, and hence can not
generate different masses for the up and down-type quarks. That
splitting must be generated on the UV brane, which is the only place
where the custodial symmetry is broken. It was demonstrated in
\cite{Csaki:2003sh} that the simplest way to do this with a complex
fermion is to introduce brane-localized fermions that can mix with
the bulk states. So we include a contribution to the brane action
for each SM right-handed fermion
\begin{gather}
 S_{\rm UV} = \int d^4 x \sum_f \left( -i \bar \xi^{f}\bar\sigma^\mu
 \partial_\mu \xi^{f} -i \eta^{f}\bar\sigma^\mu
 \partial_\mu \bar\eta^{f} + F(\eta^{f}\xi^{f} +
 \bar\eta^{f}\bar\xi^{f}) + M_f R^{1/2}(\psi^f_{R}\xi +\bar\xi
 \bar\psi^f_{R})_{z=R}
 \right)\label{eq:uvmixing}
\end{gather}
where $\xi_i$ and $\eta_i$ are the brane localized states,
$\psi_{Ri}$ is the component of the bulk state that has a zero mode,
and the index $f$ runs over all SM right-handed fermions. This leads
to the boundary condition
\begin{gather}
\psi^f_{L} = 0 \qquad \chi^f_{R} = m_n M^{f2}_m/F^2 \psi^f_{R}.
\end{gather}
The fermion KK expansion will take the form\footnote{The choice of
what powers of $z$ to include in the wavefunctions is, of course,
arbitrary. This choice makes transparent on which brane the fermion
zero mode is localized.}
\begin{gather}
\chi = \sum_n g^{(n)}(z)z^{3/2}\chi^{(n)}(x), \qquad \bar\psi=\sum_n
f^{(n)}(z)z^{3/2}\bar\psi^{(n)}(x).
\end{gather}
Again, we will normalize these canonically, giving
\begin{gather}
\int \frac{dz}{z} \chi^{(n)*}(z)\chi^{(m)}(z) = \delta_{nm},
\end{gather}
and similarly for the $\psi^{(n)}$. Note that 5D fermions are
achiral, so we can always write down a mass term in the bulk
\begin{gather}
m_5^f \bar \Psi \Psi \equiv c^fk\bar\Psi_f\Psi_f.
\end{gather}
Note that the mass itself must be $Z_2$-odd, since $\bar\Psi\Psi$
is. Even in the presence of this mass term there is still a 4D zero
mode for the orbifold even components of $\Psi$ when $v = 0$. The
$c^f$ determine the shape of the wavefunction in the extra
dimension. This would-be zero mode wavefunction is
\begin{gather}
f_f^{(0)}(z) = A_f^{(0)}\left(\frac{z}{R}\right)^{c^f - 1/2}
\end{gather}
We can see that for $c>1/2$ the zero mode is localized to the Planck
brane, and for $c < 1/2$ it is localized to the TeV brane.

\section{Kaluza-Klein spectra}\label{sec:spectra}



\FIGURE[t]{
 \includegraphics[angle=-90,width=12cm]{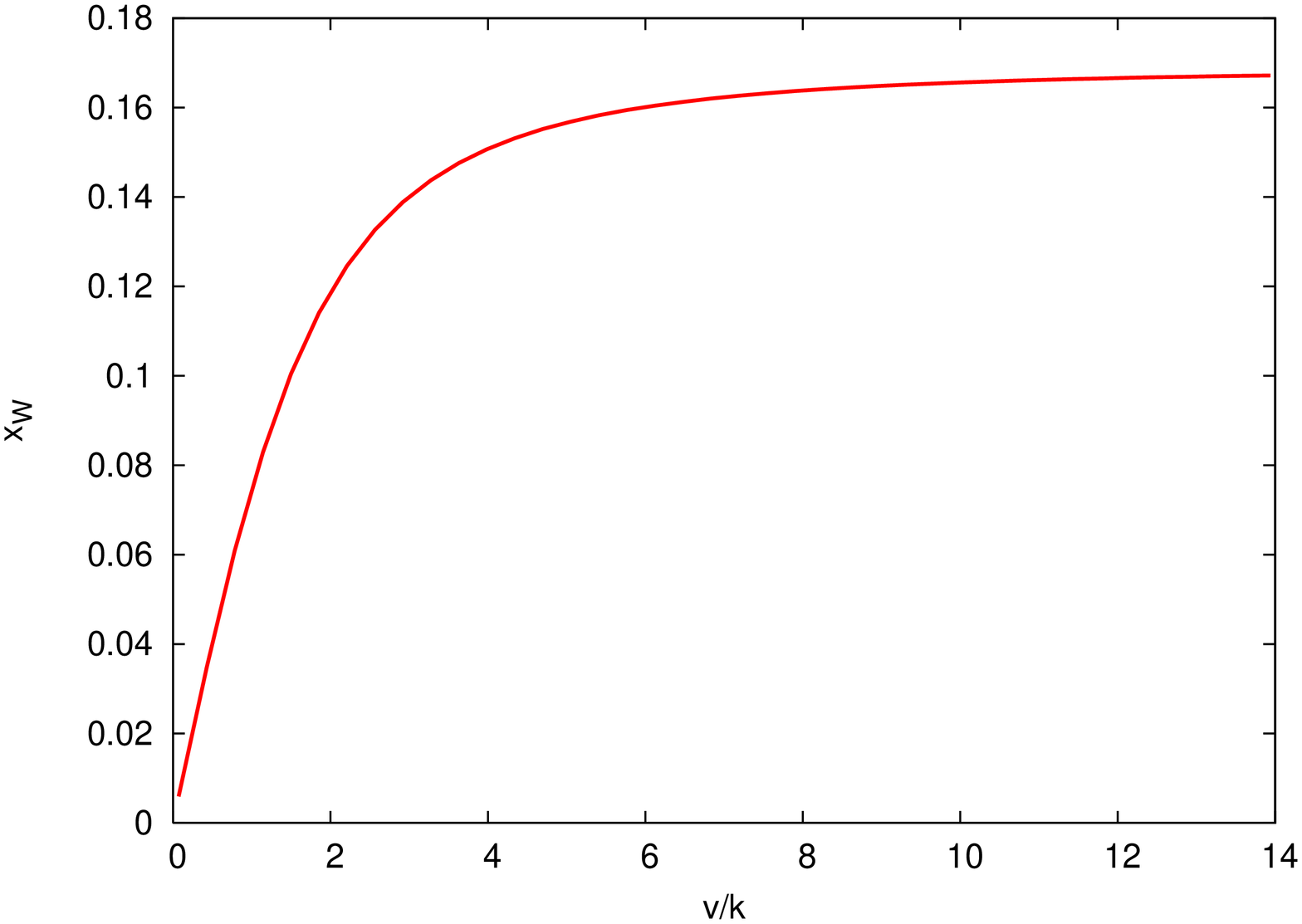}
\label{fig:wroot} \caption{Behavior of the first charged boson mass
(corresponding to the observed $W$) as a function of v/k at fixed k.
The linear behavior at small $v/k$ corresponds to the ordinary
Higgsed model limit, and the flat behavior as $v/k \to \infty$ to
the Higgsless limit.}}

We can solve Eq. (\ref{eq:gaugewf}) subject to (\ref{eq:gaugebcR})
and (\ref{eq:gaugebcRp}) to obtain $x_W^{(n)} \equiv
m_W^{(n)}/k\epsilon$. Figure \ref{fig:wroot} shows the behavior of
$x_W^{(1)}$ as a function of $v/k$. Demanding that $m_W^{(1)} =m_W$
sets the mass scale $k\epsilon$. In the region with $v/k$ small we
have $x_W^{(1)} \approx \frac{g}{2}\frac{v}{k}$ giving the standard
result $k\epsilon \approx \frac{2}{g}m_W\frac{k}{v}$. This reflects
the fact that as $v/k$ gets small, the KK scale gets large. When
$v/k$ is large, $x_W^{(1)}$ asymptotes to the Higgsless value
$x_W^{(1)2} = 1/\log(R'/R)$ \cite{Csaki:2003zu}.

\FIGURE[t]{
\includegraphics[angle=-90,width=12cm]{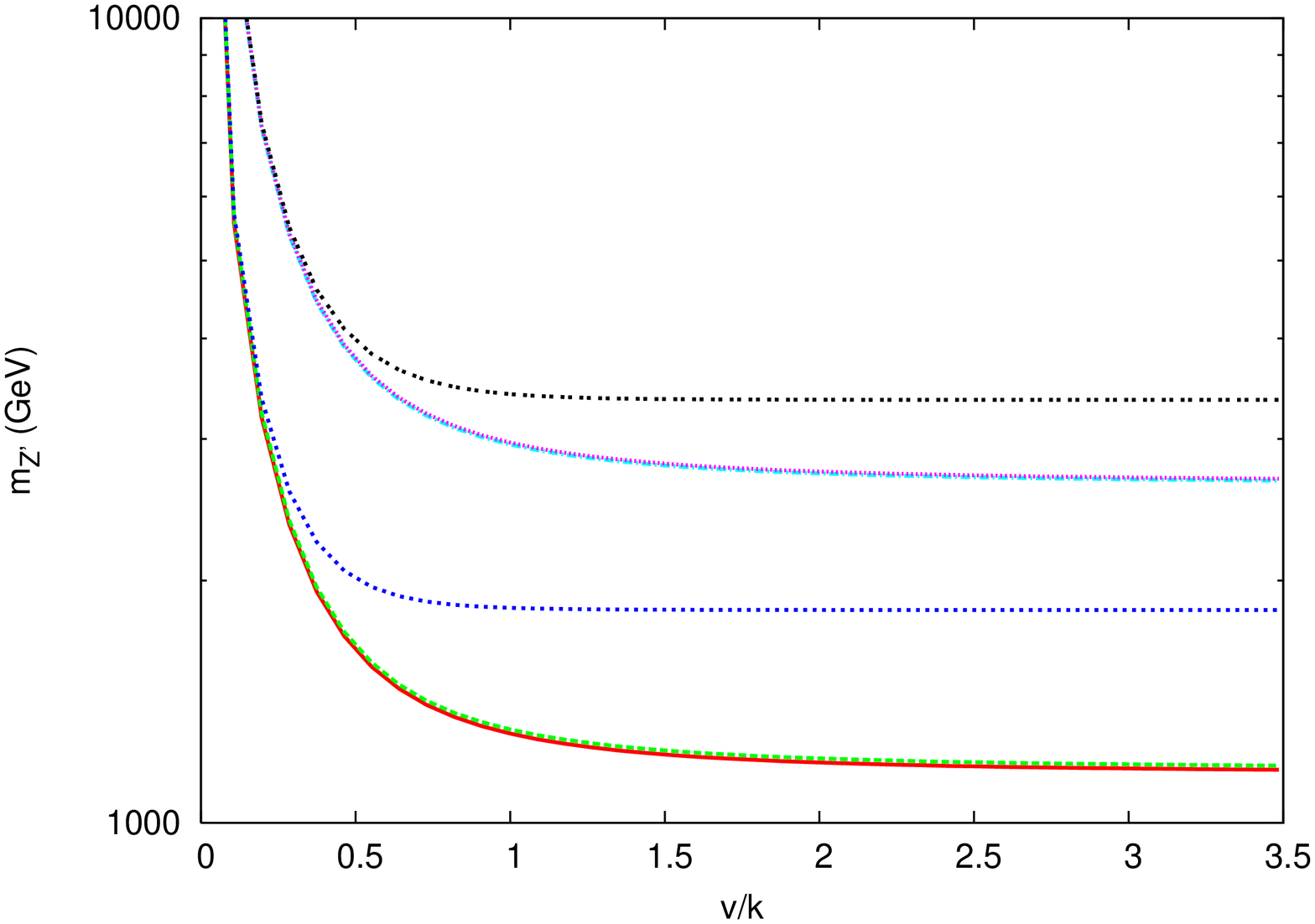}
\label{fig:zspec} \caption{Masses of the first six neutral boson KK
excitations lying above the SM $Z$ as a function of $v/k$ with the
$W$ and $Z$ masses held fixed at their physical values.}}

Once this scale has been set, we can solve for the rest of the KK
gauge boson masses. For the neutral sector this depends on the
additional parameter, $\lambda$. We can solve Eq (\ref{eq:gaugewf})
for the neutral gauge sector for $\lambda$ in terms of $x_Z^{(1)}$,
the mass of the observed $Z$ boson. We will use this to choose
$\lambda$, and hence with our choice of input parameters the
on-shell definition of the weak mixing angle, $\sin^2\theta_{\rm os}
\equiv 1 - m_W^2/m_Z^2$, is automatically set to the experimental
value $\sin^2\theta_{\rm os} = 0.222$ \cite{unknown:2003ih}. These
inputs completely determine the gauge KK mass spectrum. Figure
\ref{fig:zspec} shows this spectrum for neutral bosons. Note that
each KK level in the $v/k$ small region starts as a degenerate
triplet and splits into the doublet-singlet structure seen in
Higgsless theories as $v/k$ gets large. These masses are large
enough that these states are not bounded by direct detection
constraints at the Tevatron\cite{Abazov:2001qd}. Also, the couplings
to light fermions are small enough that they also avoid the LEP II
contact interaction constraints \cite{Davoudiasl:2004pw}.

The fermion masses depend on both the brane localized Yukawa
couplings to the Higgs and the bulk masses. Eq. (\ref{eq:fermionIR})
shows that the Yukawa couplings provide an effective Dirac mass on
the IR brane, and it is this mass that controls the lowest fermion
mass ({\it i.e.} the mass of the observed SM particle). Hence, the
relevant dimensionless parameter is $\lambda_i v/k$ (where
$\lambda_i$ is the relevant Yukawa coupling), rather than $v/k$. In
this paper we consider the case where the fermion mass hierarchy
generated by the different bulk masses, and not the Yukawas. To
correctly produce the top mass, the top/bottom Yukawa must be order
1. We will assume that the other generations have universal Yukawa
couplings $\lambda_{\rm light}$. Since we are assuming an explicit
$Z_2$ symmetry that exchanges the $SU(2)_L$ and $SU(2)_R$ we have
$c^f_L = -c^f_R \equiv c^f$ (the minus sign arises from the choice
of orbifold parities). We will pick the parameters $c^f$ to produce
the correct fermion masses for a given Yukawa coupling and Higgs
vev. In the quark sector we pick the $c^f$ to match the up-type
masses. We then introduce mixing with $UV$ brane fermions as in Eq.
(\ref{eq:uvmixing}) to generate the up-down splitting. In the lepton
sector we simply match the charged lepton masses by picking the
$c^f$, and leave the neutrinos massless. We can then solve for the
KK masses, shown in Fig. \ref{fig:fspec}. Note that a large Dirac
mass on the IR brane makes the first KK excitation light. This
behavior corresponds to that seen in \cite{Agashe:2004bm}. To avoid
light KK leptons we will need $y_{\rm light} \lesssim 1/2$. The
values of the $c^f$ depend on the value of $v/k$, but not strongly.
As an example point, at $v/k \sim 1/10$ we have $c^t = 0.43$, $c^c =
0.53$, $c^u = 0.63$, $c^\tau = 0.52$, $c^\mu = 0.58$, and $c^e =
0.66$. In the Higgsless regime these are all shifted up by about
$0.03$.

\FIGURE[t]{
\includegraphics[angle=-90,width=12cm]{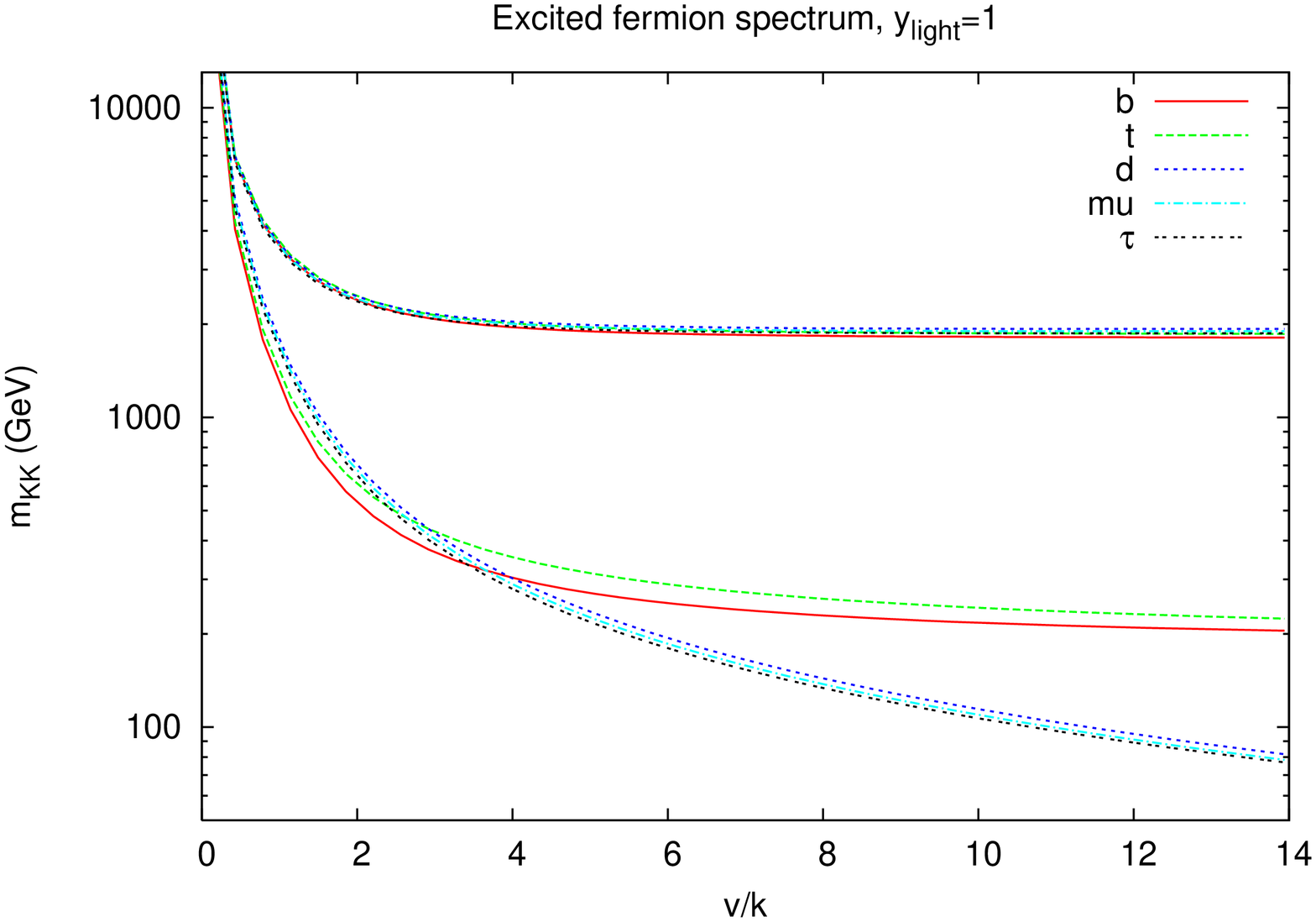}
\includegraphics[angle=-90,width=12cm]{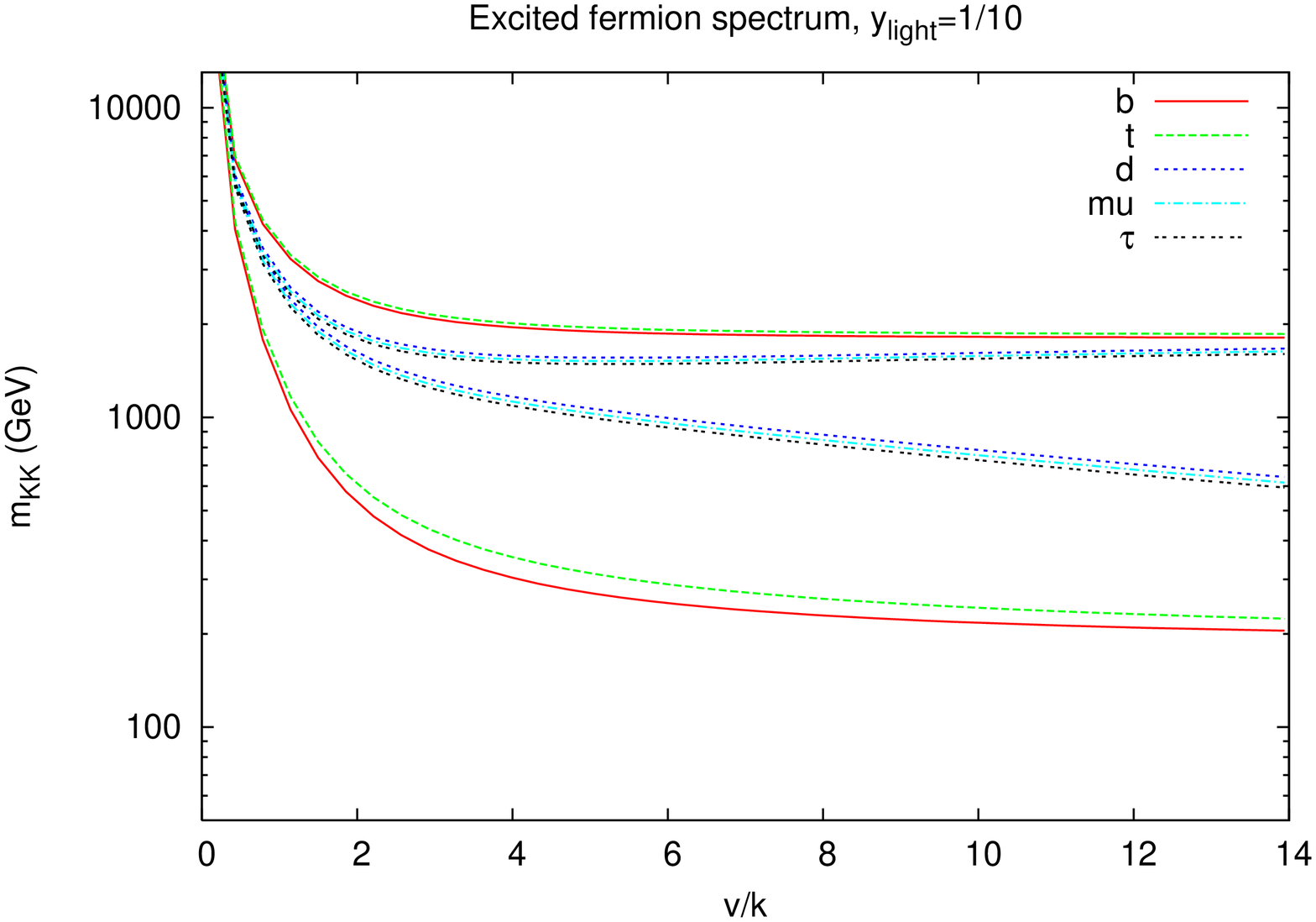}
\label{fig:fspec} \caption{Masses of the first two excited KK
fermions as a function of $v/k$ for several species. Note that the
mass of the first excitation depends strongly on the IR-brane mass
term, but the second excitation does not.}}

\section{Gauge-fermion couplings}\label{sec:gaugeff}

Here we will examine how the shifts in couplings of the $W$ and $Z$
to fermions depend on the Higgs vev. The 5D covariant derivative
acting on a fermion $\psi$ is (suppressing Lorentz indices)
\begin{gather}
D \psi = \left( \partial_M + \int_R^{R'}\frac{dz}{z}g_{5L}\left(
 A_{L\mu} T_L + \kappa A_{R\mu} T_R + \lambda Q_{B-L}A_{B\mu}\right)\right)\psi
\end{gather}
with $Q_{B-L} = (B-L)/2$. The electromagnetic charge is $Q = T^3_L +
T^3_R + Q_{B-L}$. We can rewrite the pieces of $D$ corresponding to
the neutral gauge boson couplings as
\begin{gather}
g_{5L}(I^f_{3L} -\lambda I^f_B) \left(T^3_L + \frac{\kappa I^f_{3R}
- \lambda I^f_B}{I^f_{3L} - \lambda I^f_B}T^3_R + \frac{\lambda
I^f_B}{I^f_{3L} -\lambda I^f_B}Q\right).\label{eq:neutralgauge}
\end{gather}
Where the $I^f_i = \int_R^{R'}dz/z\, \zeta_i \bar\psi_f \psi_f$
encode the extra dimensional physics. This can be matched onto the
covariant derivative from the effective 4D theory
\begin{gather}
g_{Z_1 f\bar f}(T^3_L + \sin^2\theta_{R,f} T^3_R -
\sin^2\theta_{\rm eff,f}Q).
\end{gather}
This identifies the strength of the $Z$ coupling to fermion $f$ as
$g_{Z_1 f \bar f} = g_{5L}(I^3_L - \lambda I_B)$, the effective weak
mixing angle for that coupling $\sin^2\theta_{\rm eff, f} = -\lambda
I_B/(I^3_L - \lambda I_B)$, and a new quantity that measures the
strength of the right-handed couplings: $\sin^2\theta_{R,f} =
(\kappa I^3_R - \lambda I_B)/(T^3_L - \lambda I_B)$. We can also
write an expression similar to Eq. (\ref{eq:neutralgauge}) for the
charged sector
\begin{gather}
g_{5L}I^f_{\pm L} \left(T^\pm_L + \frac{\kappa I^f_{\pm R}}{I^f_{\pm
L}} T^\pm_R\right),
\end{gather}
giving the strength of the left and right handed couplings.

Note that the wavefunctions for all electroweak particles, with the
single exception of the zero mode photon, have non-trivial
dependence on the extra dimension. In particular the different
flavors of fermions will have different wavefunctions that can be
probed by the $W$ and $Z$. Hence, all quantities defined above will
depend on the fermion species, as indicated by the label $f$.

Since the photon wavefunction is flat in the extra dimension, the
electromagnetic coupling is simply given by (using the normalization
from Eq. (\ref{eq:neutralnorm}))
\begin{gather}
 e^2 =
 \frac{g_{5L}^2}{R\log(R'/R)}\frac{\kappa^2\lambda^2}{\kappa^2+\lambda^2+
 \kappa^2\lambda^2}.
\end{gather}
We can use this to define the 5D coupling in terms of the fine
structure constant $\alpha$ by
\begin{gather}
 \frac{g_{5L}^2}{R} = 4\pi\alpha \log(R'/R)\left(
 1+\frac{1}{\lambda^2}+\frac{1}{\kappa^2}
 \right).
\end{gather}
The advantage of this definition is that it is the only coupling
that is independent of the fermion species, and allows us to relate
$g_{5L}$ to the measured quantity $\alpha$ without ambiguity.

\FIGURE[t]{
\includegraphics[angle=-90,width=12cm]{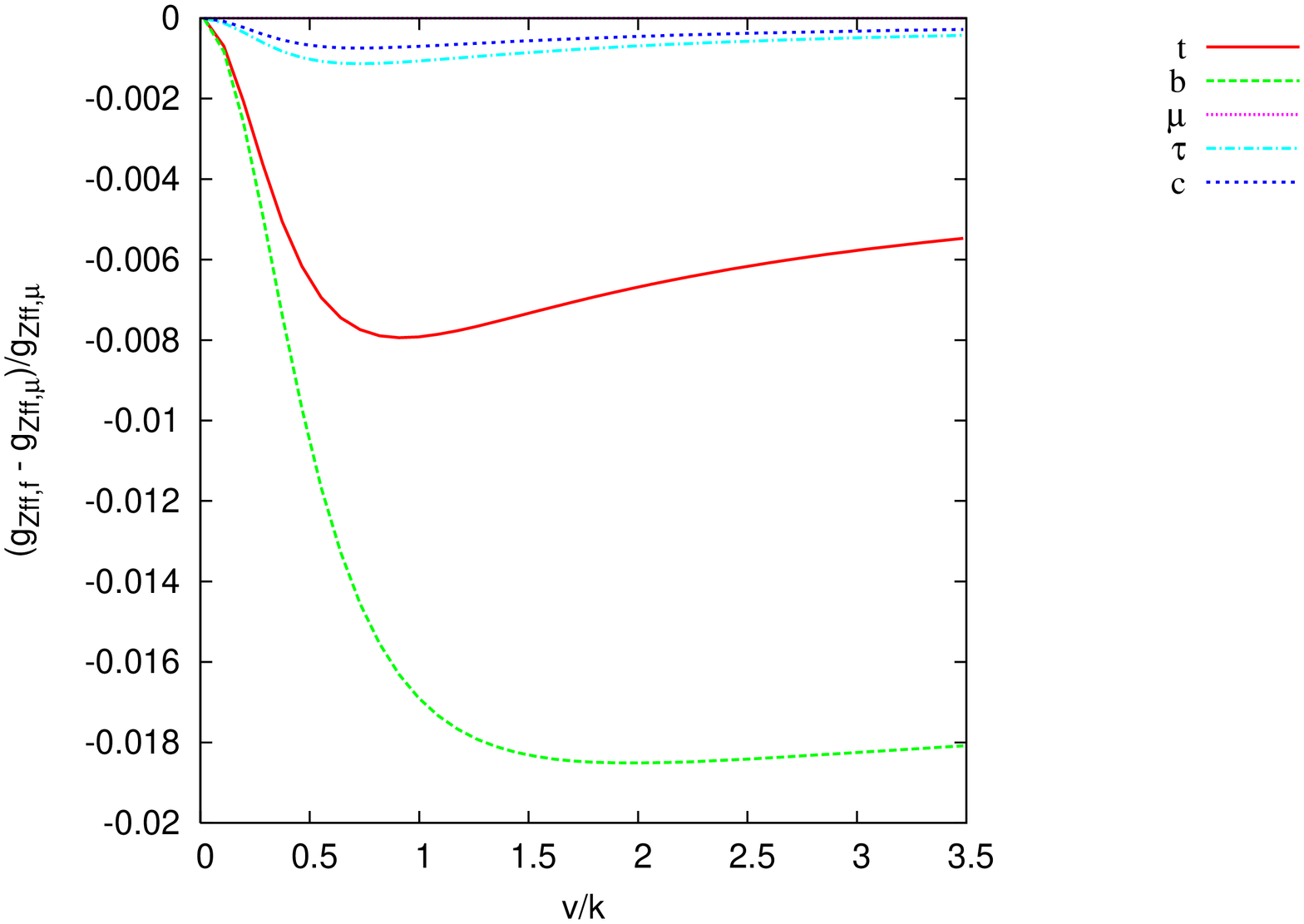}
\label{fig:zff} \caption{Shifts in the coupling of fermions to the
$Z$ induced by the $Z$ wavefunction distortions. The large shift to
the $Zb\bar b$ coupling is the dominant constraint on $v/k$.}}
With these definitions we can find the shifts in the couplings. Fig
\ref{fig:zff} shows the shift in $g_{Zf\bar f}$ as a function of
$v/k$. Note that the shifts are only large for the third generation
quarks. This is expected since they are the only fermions with
substantial overlap on the IR brane where the $W$ and $Z$
wavefunctions are distorted. The difference in effects between the
top and bottom is due to the $SU(2)_R$ violating mixings on the
Planck brane, which separately distort the $t$ and $b$
wavefunctions. Imposing the LEP and SLD bound on the shift in the
$Zb\bar b$ vertex of $\sim 1\%$ \cite{unknown:2003ih}, we find that
$v/k < 1/2$. This gives $k\epsilon > 800 \gev$, and $v\epsilon < 400
\gev$, and hence KK masses of roughly $2$ TeV. This constraint is
weaker than that quoted in \cite{Agashe:2003zs}. The discrepancy can
be traced to the different values of $c^t$ used, which is due to the
allowance here of a larger Lagrangian level Yukawa coupling ($\sim
5$), near the purturbativity bound (that the 4D coupling not exceed
$4\pi$). If we reduce the top Yukawa coupling and $c^t$ to that in
\cite{Agashe:2003zs}, the bound from the $Zb\bar b$ coupling would
shift to about $v/k \le 1/4$, corresponding to a KK mass $\sim 4$
TeV, in agreement with \cite{Agashe:2003zs}. It turns out that the
only Higgs properties that follow that is affected by this
difference are the corrections to $gg \rightarrow h$ and $h
\rightarrow \gamma\gamma$. Since they only depend on this difference
at the $20 \%$ level, this point is not crucial, and one could
simply adopt the stronger bound.
\FIGURE[t]{
\includegraphics[angle=-90,width=12cm]{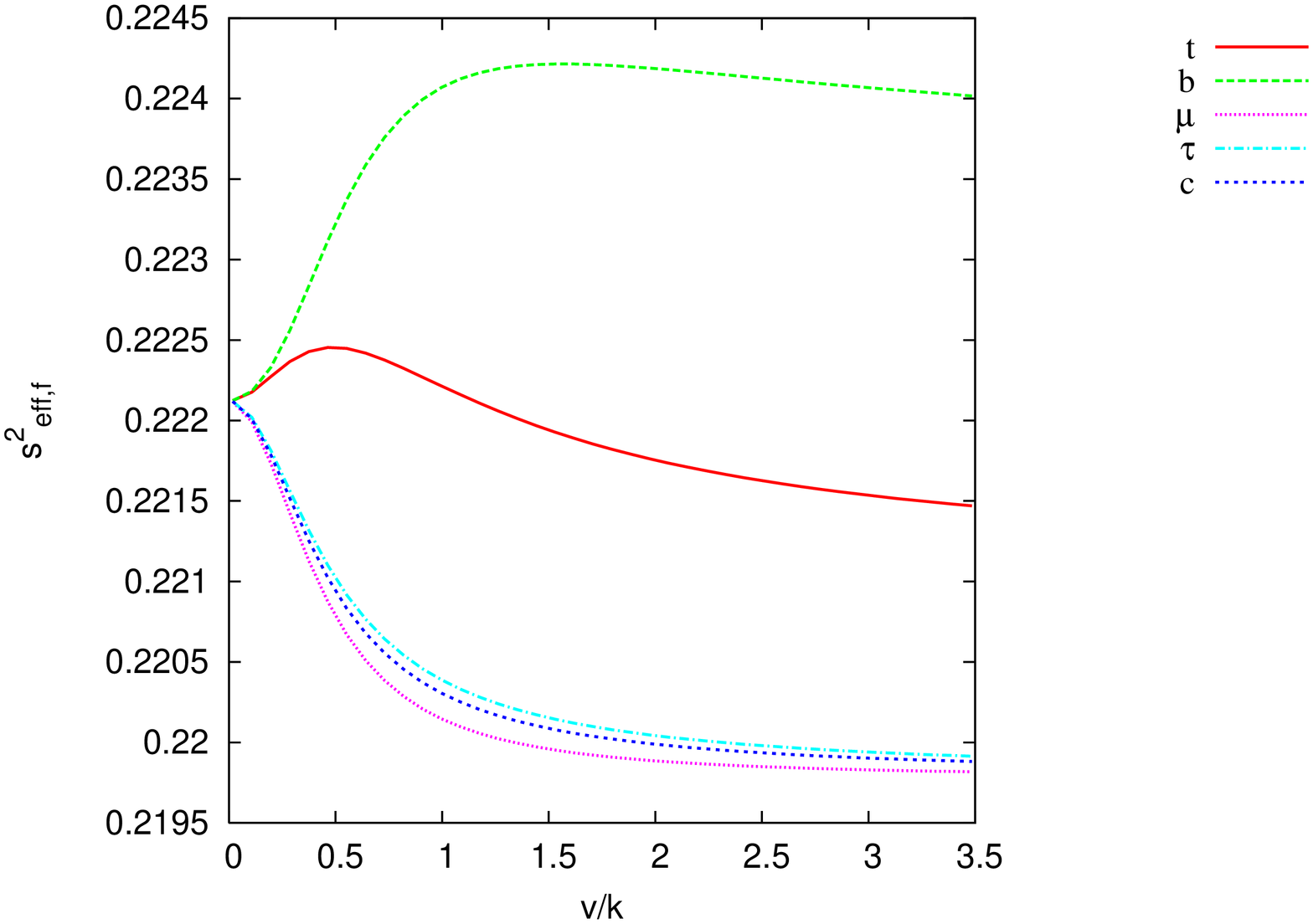}
\label{fig:seff} \caption{Corrections to the effective weak mixing
angle for couplings of fermions to the $Z$ on the $Z$-pole.}}
In Fig. \ref{fig:seff} we see the shifts in the effective weak
mixing angle in $Z$-pole observables relative to the on-shell value.
The experimental error on this measurement is $\pm 0.00036$
\cite{unknown:2003ih}, so for $v/k \le 1/4$ the model shifts
correspond to a $2\sigma$ deviation.
\FIGURE[t]{
\includegraphics[angle=-90,width=12cm]{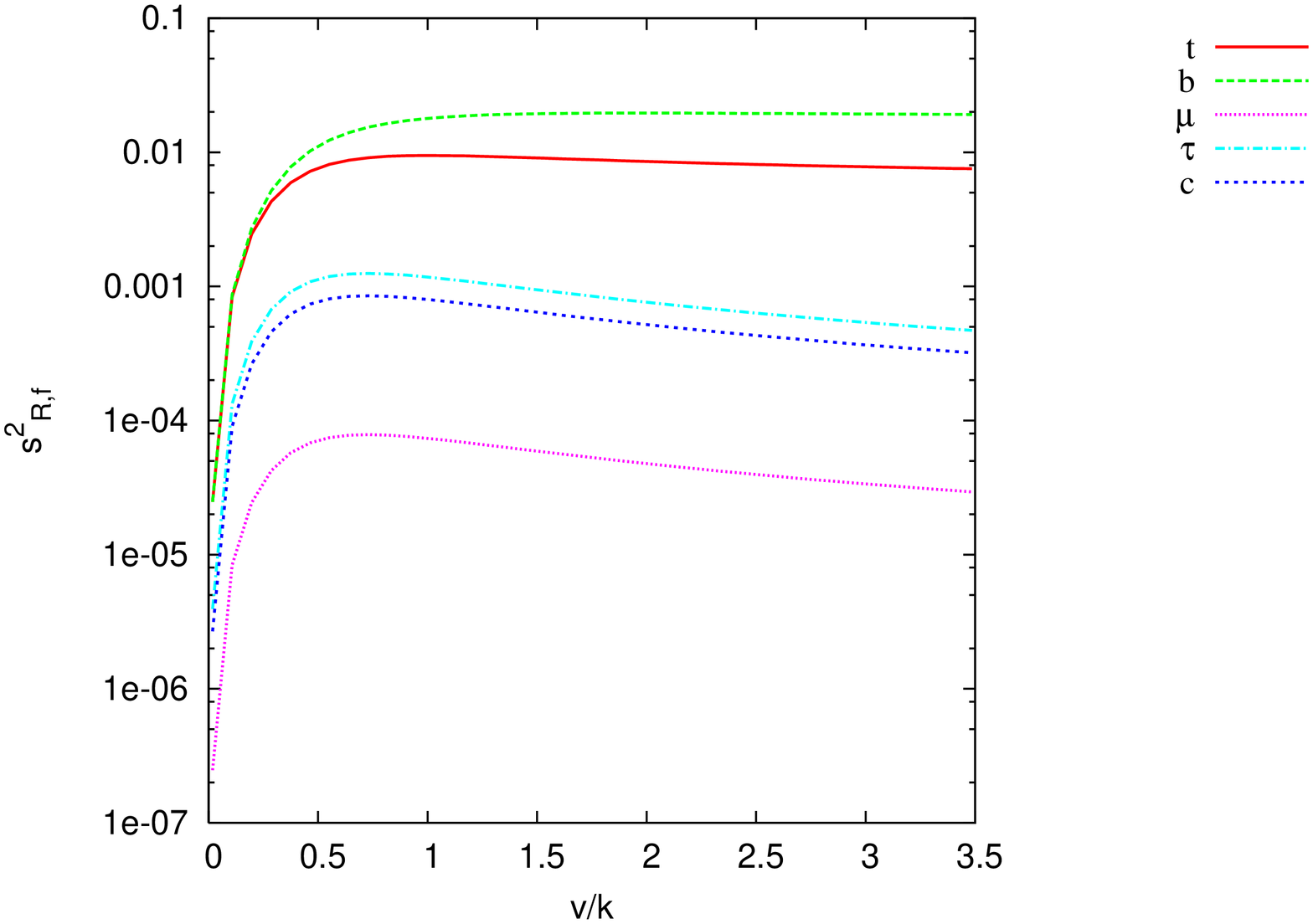}
\label{fig:sr} \caption{The effective size of the right-handed
currents induced by the $SU(2)_R$ gauge bosons.}}
Fig. \ref{fig:sr} shows $\sin^2\theta_R$, the magnitude of the right
handed couplings to SM fermions relative to the left handed
coupling. Note that the boundary conditions in Eq. \ref{eq:gaugebcR}
cause this to vanish for Planck brane localized fermions. Indeed we
see that the closer the fermion to the UV brane, the smaller the
effect. In all cases, however, the effect is unobservable in the
allowed region $v/k \le 1/4$.

\section{Higgs couplings}\label{sec:higgs}

We now investigate the shifts in couplings of particles to the
physical Higgs boson. As shown above, the main effects come from
distortions of the gauge wavefunctions near the IR brane.
Schematically, the coupling of a Higgs to two bulk modes will simply
be the product of the bulk wavefunctions evaluated at the IR brane,
times a Lagrangian parameter. So for the coupling to gauge bosons we
have
\FIGURE[t]{
\includegraphics[angle=-90,width=12cm]{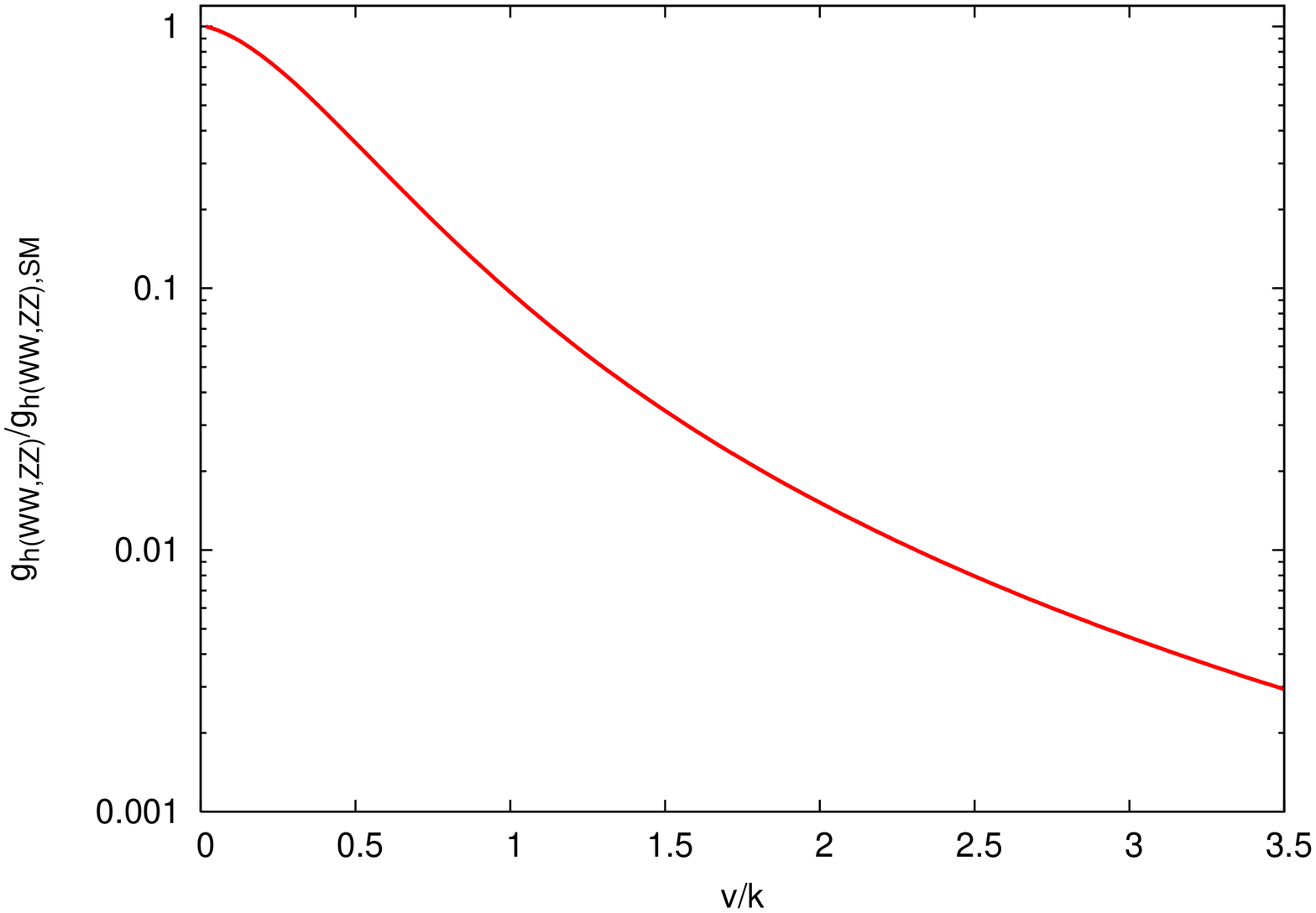}
\label{fig:wcoup} \caption{Coupling of the Higgs to vector boson
pairs compared to the SM value as a function of $v/k$. Again, the
$W$ and $Z$ coupling strengths are nearly identical due to the
custodial symmetry.}}
\begin{gather}
 g_{hWW} = \frac{g_{5L}v}{2m_W}\frac{1}{1+\kappa^2}
 \left(
 \zeta^{(1)}_{A^\pm_L}(R')-\kappa\zeta^{(1)}_{A^\pm_R}(R')
 \right)^2,
\end{gather}
and
\begin{gather}
 g_{hZZ} = \frac{g_{5L}v}{2m_Z}\frac{1}{1+\kappa^2}
 \left(
 \zeta^{(1)}_{A^3_L}(R')-\kappa\zeta^{(1)}_{A^3_R}(R')
 \right)^2.
\end{gather}
The wavefunction suppression near the TeV brane will decrease these
couplings, with $g_{h(WW,ZZ)} \to 0$ as $v/k \to \infty$. This
coupling is shown in Fig. \ref{fig:wcoup}.\footnote{Note that the
constraint $v/k \le 1/4$, coming from precision electroweak
observables, is highly sensitive to variations on the model. Many of
the features discussed in this section, however, are generic.
Consequently, we will continue to examine the full range of $v/k$,
keeping in mind the electroweak constraints.} This reduction will
weaken the LEP bound on the Higgs mass \cite{Barate:2003sz}. For
$v/k \le 1/4$ the constraint is relaxed only a few GeV. However, the
bound is moves rapidly after that point, so for $v/k = 1/2$ the
bound is $80 \gev$.

Using the above relations we can find the width for the decay into
vector pairs, which is simply
\begin{gather}
\Gamma(h\rightarrow (WW,ZZ)) =\left(\frac{1}{2}\right)
\frac{g^2_{h(WW,ZZ)}m_{(W,Z)}^3}{64 \pi
m_h^2}(\xi-4)^{1/2}(12-4\xi+\xi^2),\label{eq:realgauge}
\end{gather}
with $\xi = m_h^2/m_{W,Z}^2$, and the first factor of $1/2$ is a
symmetry factor relevant only for the $ZZ$ final state.

The decay modes where one vector is off-shell can also be important
\cite{Rizzo:1980gz}. These decay widths are
\begin{align}
 \Gamma(h\rightarrow WW^*) & = \frac{3g^2_{hWW}g^2
 m_H}{512\pi^3}H_W(m_W/m_H),\\
 \Gamma(h\rightarrow ZZ^*) & =\frac{g^2_{hZZ}g^2
 m_H}{2048\pi^3\cos^4\theta}\left(7- \frac{40}{3}\sin^2\theta_W +
 \frac{160}{9}\sin^4\theta_W\right)H_Z(m_Z/m_H),
\end{align}
where
\begin{gather}
 H_{W,Z}(x) = \int_{2x}^{1+x^2} dy
 \frac{(y^2-4x^2)^{1/2}}{(1-y^2)^2+x^4 \Gamma^2_{W,Z}/M_{W,Z}^2}
 (y^2-12x^2y+8x^2+12x^4),
\end{gather}
and we have ignored the corrections to the $Wf\bar b$ and $Z f\bar
f$ couplings which are small for $v/k \le 1/4$. It is necessary to
include the effects of the finite widths, $\Gamma_{W,Z}$, to match
onto Eq. (\ref{eq:realgauge}).

The fermion couplings to the Higgs are similar; they take the form
of the fermion wavefunctions evaluated on the TeV brane times a
Yukawa coupling. Specifically,
\begin{gather}
\lambda_{f,n} = \lambda_f \left(\chi^{f(n)}_L(R')\psi^{f(n)}_R(R') -
\psi^{f(n)}_L(R')\chi^{f(n)}_R(R')\right).
\end{gather}
Note that, since the Kaluza-Klein excitations are localized near the
TeV brane, this coupling will be enhanced by the factor
$\sqrt{\log(R'/R)}$. In the case of the 3rd generation quarks, which
have ${\mathcal O}(1)$ Lagrangian level Yukawa couplings, these
enhancements make the couplings quite large, though in all cases
they are less than $4\pi$, and hence perturbative. (Of course, with
large couplings the higher order effects are likely to be important,
but for the purposes of this paper the rough size indicated by the
tree-level result is sufficient). The width into fermion pairs is
simply
\begin{gather}
\Gamma(h \to f\bar f) = \frac{N_c
\lambda^2_{f,1}m_h}{32\pi}\left(1-\frac{4
m_f^2}{m_h^2}\right)^{3/2},
\end{gather}
where $N_c$ counts the fermion's color degrees of freedom.

Finally, two of the most important couplings for the discovery of
the Higgs boson at the LHC are the Higgs-glue-glue, and
Higgs-gamma-gamma vertices. These couplings are absent at
tree-level, but are generated radiatively by loops containing
fermions and $W$-bosons. In the present model, these vertices
receive corrections from two sources. First, the KK excitations of
all fermion species can run in the loop (along with the $W$ KK
excitations in the case of the $\gamma\gamma$ couplings); since
these have substantial couplings to the Higgs these corrections can
be large. Second, the suppression in $g_{HWW}$ from the distortion
of the $W$ wavefunction can suppress the coupling of the Higgs to
two photons. We can calculate in the effective 4D theory obtained by
doing the Kaluza-Klein reduction. After doing this, all of the
information about the extra dimensional physics is contained in the
masses and couplings of the KK states. Hence, to calculate these
contributions, we can adapt the formulae from
\cite{Petriello:2002uu} (see also \cite{Gunion:1989we}), which
computed these shifts in the context of Universal Extra Dimensions,
by simply replacing the masses and couplings.
\FIGURE[t]{
\includegraphics[angle=-90,width=12cm]{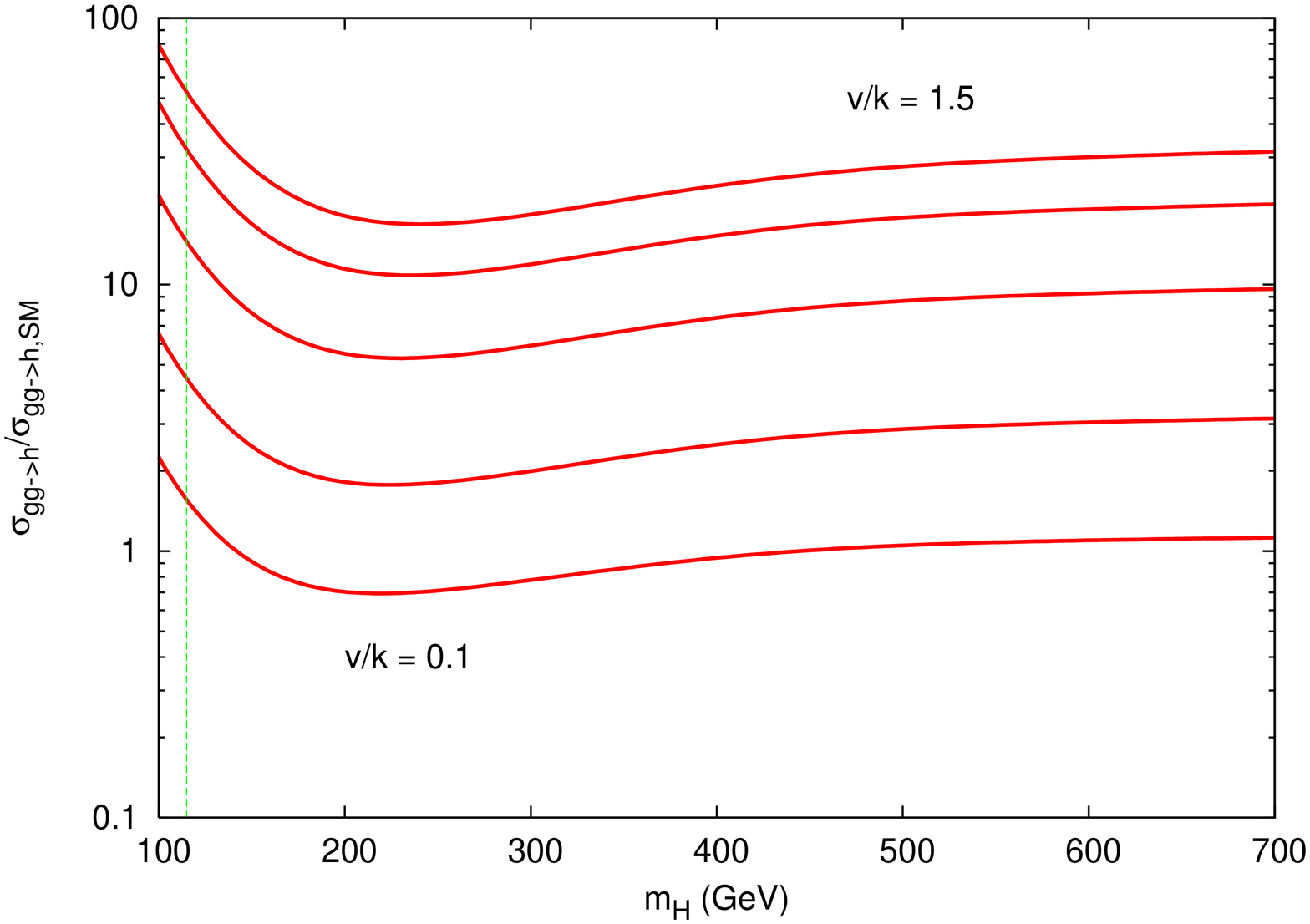}
\label{fig:hgg} \caption{Ratio (at lowest order) of the production
cross section for $gg\to h$ compared with the value in the SM. The
different curves correspond to the values of $v/k$ from top to
bottom of (1.5, 1.1, 0.8, 0.4, 0.1). Here we have taken $y_{\rm
light} = 1/10$.}}
The parton level cross section for producing a Higgs from gluon
fusion is
\begin{gather}
\sigma_{gg\to h} = \frac{G_F[\alpha_s(m_H)]^2}{32\sqrt{2}\pi
m_H^4}\left| \sum_i F^i\right|^2.
\end{gather}
Here the sum runs over all KK states (including the zero modes) of
all colored fermions, with $i$ labeling the flavor. The kinematic
function $F^i$ is
\begin{gather}
F^i =
2m_i\sum_{n}m_{i,n}\lambda_{i,n}\{-2+(m_H^2-4m_{i,n}^2)C_0(m_{i,n})\},\label{eq:ft}
\end{gather}
where $C_0(x)$ is an abbreviation for the three-point scalar
Passarino-Veltman function \cite{Passarino:1978jh}
\begin{gather}
C_0(x) = C_0(m_H^2,0,0;x,x,x).
\end{gather}
This can be expressed simply as \cite{Petriello:2002uu}
\begin{gather}
 C_0(m^2) = \left\{ \begin{array}{r@{\quad:\quad}l}
 -\frac{2}{m_H^2}\left[
 \arcsin\left(\frac{1}{\sqrt{\tau}}\right)\right]^2 & \tau \ge 1 \\
 \frac{1}{2m_H^2}\left[\ln
 \left(\frac{1+\sqrt{1-\tau}}{1-\sqrt{1-\tau}}\right)\right]^2 &
 \tau < 1
 \end{array} \right. ,
\end{gather}
with $\tau = 4m^2/m_H^2$. Figure \ref{fig:hgg} shows the ratio of
the $gg \to h$ cross-section to that of the Standard Model as a
function of the Higgs mass for $y_{\rm light} = 1/10$. As expected,
there can be large corrections, even in the small $v/k$ region.

For the decay $h \to \gamma \gamma$ we have
\begin{gather}
\Gamma_{h\to \gamma\gamma} = \frac{\alpha^2}{\pi^3m_H
m_W^2}\left|\sum_i F^i\right|^2
\end{gather}
where the $F^i$ are as in Eq. (\ref{eq:ft}) for fermions and the sum
over $i$ now includes a contribution from the $W$
\begin{gather}
F^W_{\rm SM} = \frac{g_{hWW}}{g_{gWW,SM}}\left(
\frac{1}{2}m_H^2 + 3 m_W^2 - 3m_W^2(m_H^2-2m_W^2)C_0(m_W^2))
\right),
\end{gather}
There are also contributions from the KK modes of the $W$. For those we can not simply use the formulae from \cite{Petriello:2002uu}, since the Universal Extra Dimension setup considered there contains contributions from KK modes of the Goldstone bosons. Those modes are absent here, so that contribution must be re-calculated. We find
\begin{gather}
 F^W = \sum_n \frac{g_{h W^{(n)}W^{(n)}}}{g_{hWW,SM}}
 \left(\frac{m_H^2}{2} + 4m_W^2 -
 [3m_W^2(m_H^2-2m_{W^{(n)}}^2)-2m_H^2m_{W^{(n)}}^2]C_0(m_W^2)
 \right),
\end{gather}
\FIGURE[t]{
\includegraphics[angle=-90,width=12cm]{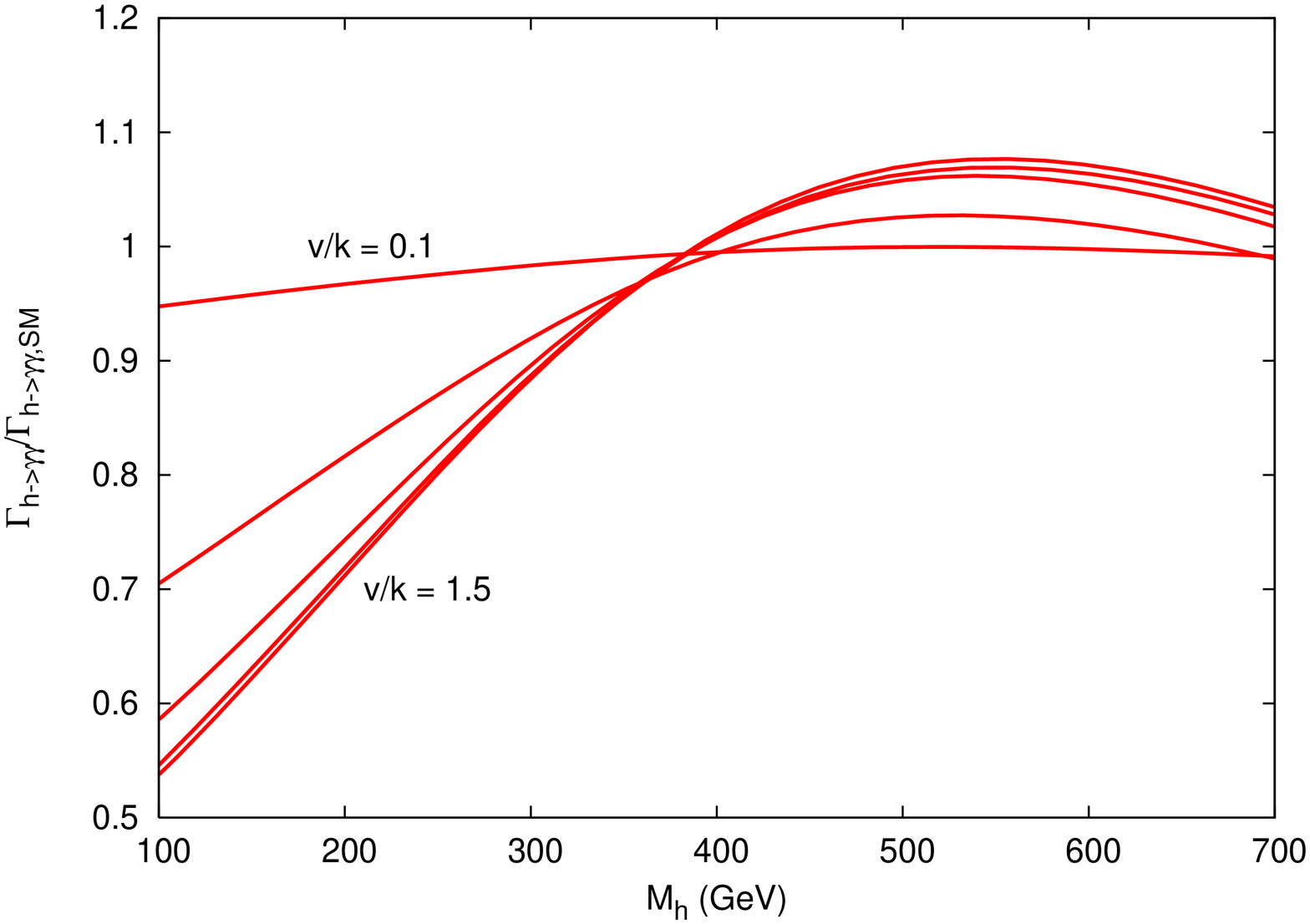}
\label{fig:hgammagamma} \caption{Ratio of the decay of a Higgs to
$\gamma\gamma$ compared with the value in the SM. The different
curves correspond to the values of $v/k$ from bottom to top on the
left edge of the graph of (1.5, 1.1, 0.8, 0.4, 0.1). Again, we have
taken $y_{\rm light} = 1/10$.}}
This decay mode is shown in Fig. \ref{fig:hgammagamma}. Note that there is some suppression for low Higgs masses.

Putting all of this together we can compute the branching ratios,
shown in Fig \ref{fig:widthbr} for $v/k = 1/10$, $y_{\rm light} =
1/10$. Several features are visible. First, the $WW$ and $ZZ$
coupling suppression results in a delayed dominance of these modes,
although they do dominate eventually for all allowed values of
$v/k$. Second, the $Z_2$ left-right symmetry requires the $b$ and
$t$ couplings to be equal, and hence the width to $b\bar b$ is
always larger than the width to $t\bar t$ (they would, of course, be
equal if $m_t = m_b$). Finally, this enhancement in the $b$ coupling
suppresses all other modes. In particular, the $h\to \gamma\gamma$
mode is unobservable. Note that even if the $b$ coupling were not
enhanced, the $\gamma\gamma$ mode would be reduced over a large
region of parameter space by the $hWW$ coupling suppression. This
means searches for Higgs bosons at the LHC which depend on the $h\to
\gamma\gamma$ decay mode will have a reduced signal, and
will possibly not be viable if this model is correct.

\FIGURE[t]{
\includegraphics[angle=-90,width=12cm]{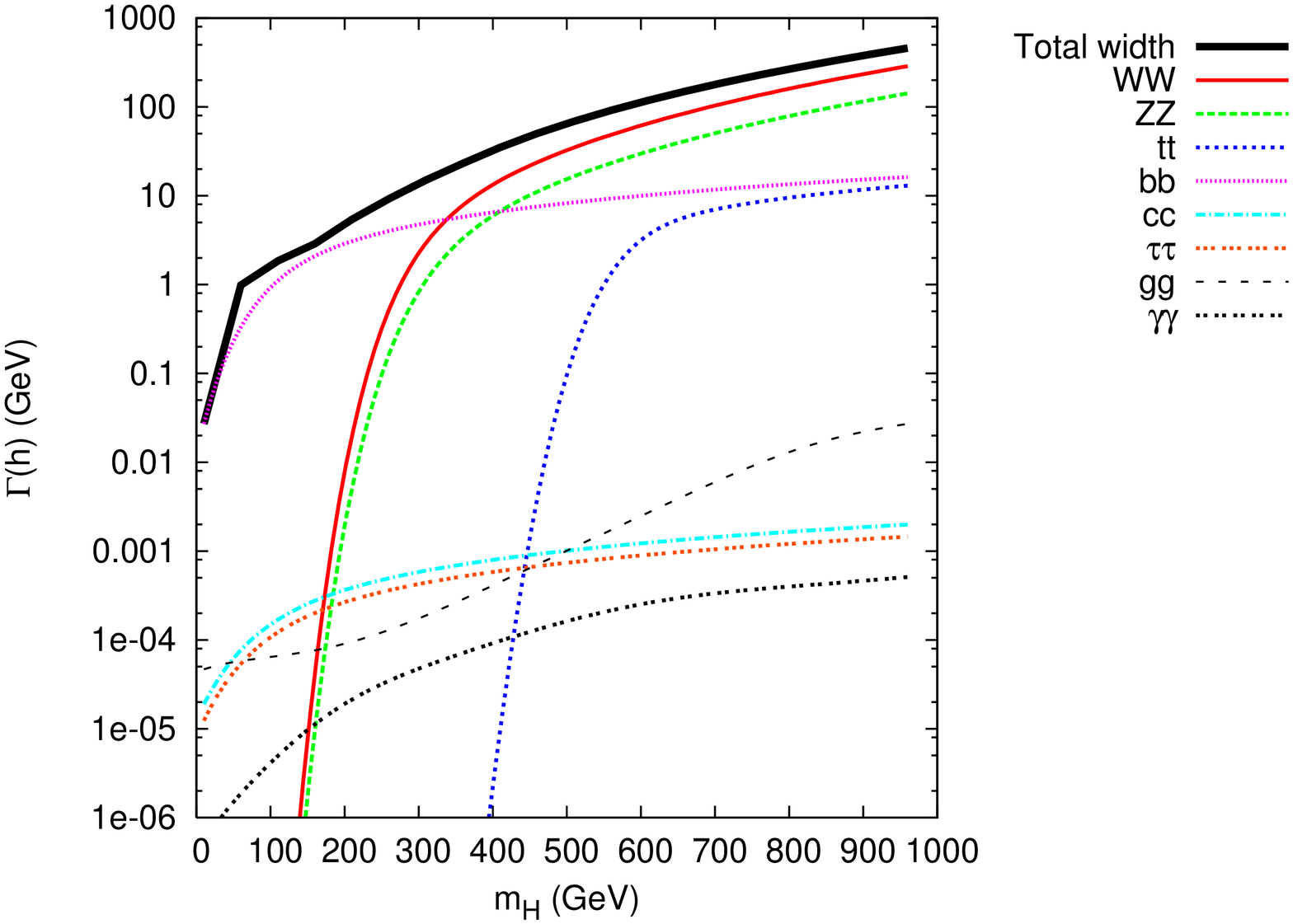}
\includegraphics[angle=-90,width=12cm]{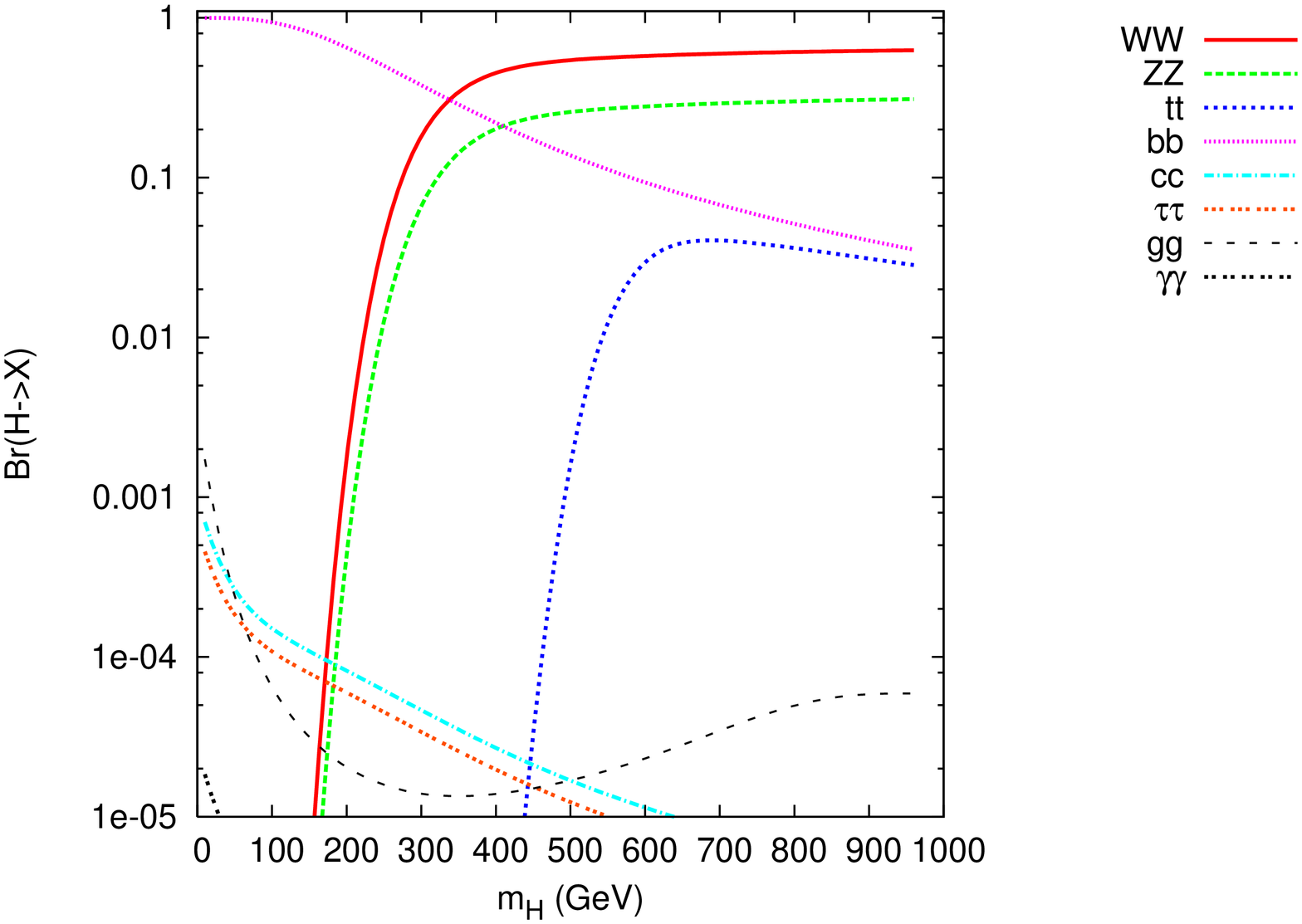}
\label{fig:widthbr} \caption{Partial widths (top) and branching
ratios (bottom) for Higgs decay into various channels as a function
of the Higgs mass at fixed $v/k = 1/10$ and $y_{\rm light =
1/10}$.}}

It has been observed that Higgsless models with KK masses $\gtrsim 1
\tev$ show a breakdown of perturbative unitarity in longitudinal
gauge boson scattering \cite{Davoudiasl:2003me}. Of course, in a
purely Higgsed 4D model, unitarity is maintained to arbitrarily high
scales. It is therefore interesting to see the behavior of the
amplitude for $W_L W_L \to W_L W_L$ \cite{Duncan:1985vj} as a
function of $v/k$. This is shown in Fig. \ref{fig:puvscaleh} for a
fixed Higgs mass of $150 \gev$. Note both the rapid falloff and the
fact that, in the region $v/k \le 1/4$, unitarity is maintained to
scales above 6 TeV.

\FIGURE[t]{
\includegraphics[angle=-90,width=12cm]{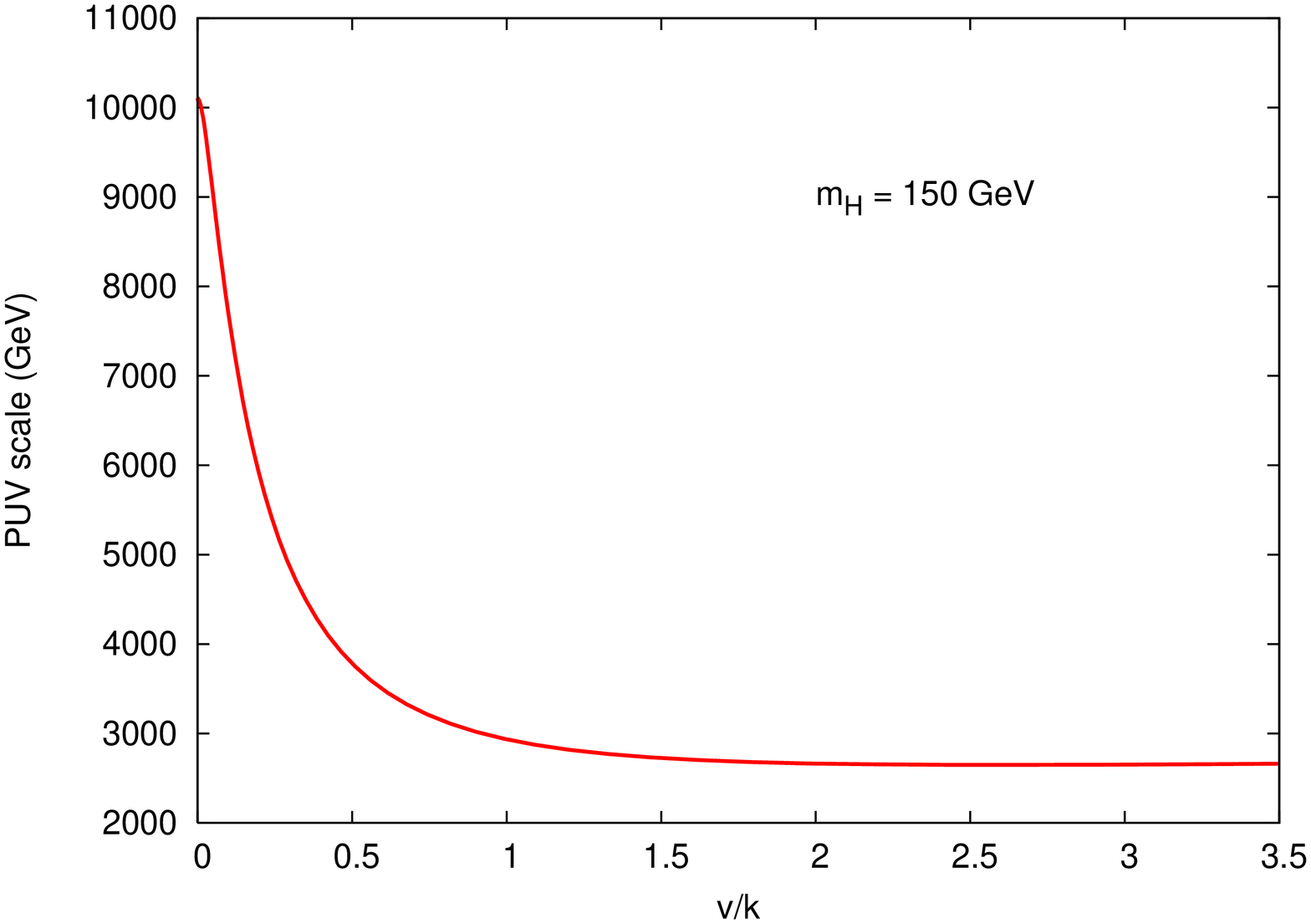}
\label{fig:puvscaleh} \caption{The center-of-mass energy at which
tree-level perturbative unitarity is violated in $W^+_L W^-_L \to
W^+_L W^-_L$ scattering.}}

\section{Conclusion}\label{sec:conclusion}

In this paper we have investigated the effects of a finite Higgs vev
in the Left-Right symmetric Randall-Sundrum model with fermions and
gauge bosons in the bulk. The main effects come from distortions of
the $W$ and $Z$ wavefunctions near the IR brane. We found that the
model is free of existing constraints as long as $v/k \lesssim 1/4$.
In this region the Higgs coupling to the gauge bosons can be
suppressed by a factor of up to $1/3$, and the Higgs couplings to
$gg$ and $\gamma\gamma$ can be substantially shifted. This results
in a new pattern of branching ratios as a function of $m_H$.

It has been shown previously that the precision electroweak
observables can be shifted by inclusion of brane localized kinetic
terms for the gauge bosons and fermions
\cite{Davoudiasl:2002ua,Carena:2002me,Carena:2004zn}. This will
shift the allowed region of $v/k$, but will not qualitatively alter
the properties of the Higgs couplings. Additionally, it would be
interesting to consider the effects of non-zero Higgs-radion mixing,
which has been set to zero here.

In this model it may be difficult to discover the Higgs, since the
$\gamma\gamma$ mode is invisible over much of the parameter space
and the massive gauge boson couplings to the Higgs are reduced.
However, when it is found the properties of the Higgs will be an
important tool in mapping out the parameters of the full model.

\acknowledgments

The author would like to thank Kaustubh Agashe, JoAnne Hewett,  Jay
Hubisz, Frank Petriello, Tom Rizzo, and Marc Schreiber for helpful
discussions.

\appendix

\section{Bidoublet Higgs}

Instead of the real bidoublet used in Section \ref{sec:formalism},
we could use a complex bidoublet Higgs field, producing a version of
the Two Higgs Doublet Model. We can parameterize this field as
\begin{gather}
\varphi = \begin{pmatrix}
  \varphi_1^0 & \varphi_1^+ \\
  \varphi_2^- & \varphi_2^0 \\
\end{pmatrix}
\end{gather}
The most general form of the potential for complex bidoublet field
is \cite{Gunion:1989in}
\begin{align}
\notag V(\varphi)&= -\mu^2\Tr\varphi^\dagger\varphi
 +\lambda_1(\Tr\varphi^\dagger\varphi)^2
 +\lambda_2\Tr\varphi^\dagger\varphi\varphi^\dagger\varphi
 +\frac{1}{2}\lambda_3(\Tr\varphi^\dagger\tilde\varphi+
 \Tr\tilde\varphi^\dagger\varphi)^2\\
 &+\frac{1}{2}\lambda_4(\Tr\varphi^\dagger-
 \Tr\tilde\varphi^\dagger\varphi)^2
 +\lambda_5\Tr\varphi^\dagger\varphi
 \tilde\varphi^\dagger\tilde\varphi
 +\frac{1}{2}\lambda_6(\Tr\varphi^\dagger\tilde\varphi
 \varphi^\dagger\tilde\varphi
 +\Tr\tilde\varphi^\dagger\varphi\tilde\varphi^\dagger\varphi),\label{eq:pot}
\end{align}
where $\tilde\varphi = \sigma_2\varphi^*\sigma_2$, and $\sigma_2$ is
the ordinary Pauli matrix. We expect that there will be solutions
where the neutral fields acquire vevs, so we try
$\langle\varphi^0_1\rangle = v_1$, $\langle\varphi^0_2\rangle =
v_2$. Stability of this solution requires the two conditions
$\partial V/\partial v_{1,2} = 0$, where all fields are evaluated at
their vevs. These then imply
\begin{gather}
\left( { v_1}^2-{ v_2}^2 \right)\left( -4\,{ \lambda_3}-{
\lambda_5}+{ \lambda_2}-{\lambda_6} \right) = 0.
\end{gather}
There is no symmetry that can require the second factor to vanish,
so we can see that aside from a vanishingly small and unnatural
region of parameter space we have $v^2_1 = v^2_2 \equiv v^2/2$. The
second stability condition then gives
\begin{gather}
v^2 = \frac{1}{2}\,{\frac
{{\mu}^{2}}{4\,{\lambda_3}+2\,{\lambda_1}+{\lambda_2
}+{\lambda_5}+{\lambda_6}}}.
\end{gather}
Inserting this solution into Eq. (\ref{eq:pot}) we can read off that
the mass eigenstates are
\begin{align}
 h & \equiv \Re(\varphi^0_1 + \varphi^0_2)  & H & \equiv \Re(\varphi^0_1 - \varphi^0_2)\notag\\
 A & \equiv \Im(\varphi^0_1 + \varphi^0_2)  & G^0 & \equiv \Im(\varphi^0_1 - \varphi^0_2)\notag\\
 h^+ & \equiv \varphi^+_1 + \varphi^+_2  & G^+ & \equiv \varphi^+_1 -
 \varphi^+_2.
\end{align}
The $G^i$ are the would-be Goldstone fields, and hence have no mass
terms in the potential. Furthermore, the structure of the potential
requires $m_H = m_{h^+}$. However, there are enough parameters that
the masses are otherwise arbitrary. To parameterize this, we can
write
\begin{gather}
m_h^2 = \lambda_h v^2,\qquad m_H^2 = m_{h^+}^2 = \lambda_H
v^2,\qquad m_A^2 = \lambda_A v^2.
\end{gather}
CP symmetry tells us that there is no three-point vertex coupling
two gauge bosons to the CP-odd scalar. Hence, restricting ourselves
to the neutral sector, we find the analysis from the main part of
the paper goes through with minimal changes.

\bibliography{paper}

\end{document}